\documentclass[aps,prd,nofootinbib,preprint]{revtex4-1}
\usepackage{graphicx,color}
\usepackage{amsmath,amssymb}
\usepackage{multirow}
\newcommand{\lsim}{\mathrel{\mathop{\kern 0pt \rlap
  {\raise.2ex\hbox{$<$}}}
  \lower.9ex\hbox{\kern-.190em $\sim$}}}
\newcommand{\gsim}{\mathrel{\mathop{\kern 0pt \rlap
  {\raise.2ex\hbox{$>$}}}
  \lower.9ex\hbox{\kern-.190em $\sim$}}}
\newcommand{\pbi}{{\;{\rm pb}^{-1}}}
\newcommand{\fbi}{{\;{\rm fb}^{-1}}}

\newcommand{\gev}{{\;{\rm GeV}}}
\newcommand{\tev}{{\;{\rm TeV}}}
\newcommand{\tb}{\bar{t}}
\newcommand{\ttb}{t\bar{t}}

\newcommand{\sg}{{\sigma}}

\newcommand{\lm}{{\lambda}}
\newcommand{\Lm}{{\Lambda}}
\newcommand{\gm}{{\gamma}}

\newcommand{\beq}{\begin{equation}}
\newcommand{\eeq}{\end{equation}}
\newcommand{\bea}     {\begin{eqnarray}}
\newcommand{\eea}     {\end{eqnarray}}
\newcommand{\bit}     {\begin{itemize}}
\newcommand{\eit}     {\end{itemize}}
\newcommand{\no}{{\nonumber}}

\newcommand{\br}{{\rm BR}}

\newcommand{\on}{ {(1)} }
\newcommand{\tw}{ {(2)} }

\newcommand{\mgtw}{ M_{g^{(2)} }}
\newcommand{\gtw}{ g^{(2)} }

\begin{document}

\title{Direct bound on the minimal Universal Extra Dimension model\\
from  the $t\bar{t}$ resonance search at the Tevatron
}

\author{Sanghyeon Chang}
\email{sang.chang@gmail.com}
\author{Kang Young Lee}
\email{kylee14214@gmail.com}
\author{So Young Shim}
\email{soyoungshim15@gmail.com}
\author{Jeonghyeon Song}%
\email{jhsong@konkuk.ac.kr}
\affiliation{
Division of Quantum Phases \& Devises, School of Physics, 
Konkuk University,
Seoul 143-701, Korea
}
\date{\today}

\begin{abstract}
In the minimal Universal Extra Dimension (mUED) model,
the second Kaluza-Klein (KK) gluon $g^{(2)}$
has loop-induced vertices with the standard model quarks,
mediated by the first KK modes of the quark and the gluon/Higgs boson.
With a top quark pair, this vertex
is enhanced by the cooperation of
the strong coupling of a gluon and the large Yukawa coupling of a top quark,
leading to substantial branching ratio of $\br(\gtw \to t \tb) \approx 7\%-8\%$.
As the $g^{(2)}$ coupling with two gluons 
appears via dimension-6 operator,
$q\bar{q}\to g^{(2)}\to t\bar{t}$
is the golden mode for the mUED model.
Hence the best channel is the $t\bar{t}$ resonance search
in $p\bar{p}$ collisions.
The recent Tevatron data at $\sqrt{s}=1.96\tev$ with
an integrated luminosity of 8.7 fb$^{-1}$
are shown to give the first direct bound on the $g^{(2)}$ mass
above $800\gev$.
The implication and future prospect at the LHC are discussed also. 
\end{abstract}

\maketitle

\section{Introduction}
\label{sec:introduction}

%\key{Why top}
As the heaviest known fundamental particle,
the top quark has unique properties 
within the standard model (SM)~\cite{top}.
Its large Yukawa coupling to the Higgs boson 
enhances the loop-induced vertex of $h$-$g$-$g$,
which leads to sufficiently large gluon fusion production of the SM Higgs boson.
The large top quark mass also preserves its fundamental properties
since it decays promptly before the hadronization,
which enables us to measure 
the $W$ helicity in the top decay~\cite{W:helicity}
and the top spin correlations~\cite{top:spin}.
Phenomenologically more attractive is that the top quark can be tagged.
If produced near the threshold,
the top quark is identified with a $b$-tagged jet and a $W$, or three jets of which the invariant mass
is near the top quark mass.
If highly boosted, it is tagged by
the substructure of collimated jets~\cite{top:tagging}.

%\key{Why $\ttb$}
The top tagging opens new channels to probe new physics beyond the SM,
especially through resonant $\ttb$ production.
The $\ttb$ resonances have been searched 
at the Tevatron~\cite{CDF:ttb:semileptonic,D0:ttb:semileptonic,CDF:ttbar:hadronic,Tevatron:4.8}
and at the LHC~\cite{ATLAS:ttb,CMS:ttb}.
No experiment has found any significant evidence,
placing upper bounds on the production cross section times the 
branching ratio into $t\tb$.
The most recent results 
have been reported in the 36th International Conference on High Energy Physics 
(ICHEP 2012) 
based on the data with a total luminosity of $2.05\fbi$ by the ATLAS experiment~\cite{ATLAS:2.05},
$5\fbi$ by the CMS~\cite{CMS:5}, and $8.7\fbi$ by the Tevatron~\cite{Tevatron:8.7}.

%\key{NP models and Benchmark scenario}
Various new physics models have candidates for
$\ttb$ resonances
such as
CP-even and CP-odd heavy Higgs bosons in the MSSM~\cite{mssm:Higgs},
a scalar resonance in two-Higgs-doublet models~\cite{2HDM},
a vector particle like $Z'$ in extended gauge theories~\cite{Zprime},
$Z'$ in  top-color assisted technicolor model~\cite{techni},
massive color-octet gauge
bosons~\cite{color:octet},
a coloron~\cite{coloron},
the first Kaluza-Klein (KK) gluon in
the bulk Randal-Sundrum model~\cite{RS:KK:gluon}
and in one or two extra dimensional models~\cite{Rizzo:KK:gluon}.

%\key{Why UED?}
Another interesting candidate for $\ttb$ resonance
is the second KK gluon 
in the five-dimensional (5D) Universal Extra Dimension (UED) 
model\,\cite{ued}.
This model has an additional single flat extra dimension of size $R$,
compactified on an $S_1/Z_2$ orbifold.
All the SM fields propagate freely in the whole five-dimensional spacetime,
each of which has an infinite number of KK
excited states.
This model has drawn a lot of interest as providing
solutions for proton decay\,\cite{proton},
the number of fermion generations\,\cite{fermion-generation-number},
and supersymmetry breaking\,\cite{susy-break}.
Most of all, 
the conservation of KK parity makes this model more appealing:
the compactification scale $R^{-1}$ can be as low as about 300 GeV;
the lightest KK particle (LKP) is a good
candidate for the cold dark matter~\cite{KKcdm}.
In the minimal version called the mUED model,
incalculable boundary kinetic terms are assumed to
vanish at the cut-off scale $\Lm$, leading to definite and well-defined
radiative corrections to KK masses.

There are various {\em indirect} constraints on the lower bound on $R^{-1}\gsim 300\gev$
from the $\rho$ parameter \cite{rho},
the electroweak precision tests \cite{EWPT},
the muon $(g-2)$ measurement \cite{gmuon},
the flavor changing neutral currents \cite{fcnc}, 
and the recent measurement of the Higgs boson mass 125 GeV~\cite{mued:higgs}.
An upper limit on $R^{-1}\lsim 1.6\tev$ is from dark matter constraints 
to avoid overclosing the 
universe~\cite{KKcdm,Kakizaki}.
To date, however, no direct limits on the mUED have been placed by collider signatures,
despite the intensive studies~\cite{collider:pheno}.
Difficulties are generic because of nearly degenerate KK mass spectrum.
The most accessible new particles are the first KK modes produced in pairs,
each of which decays into the missing LKP and a SM particle.
Very small mass gap between any first KK mode and the LKP
results in quite soft SM particles which are very challenging to observe
especially at hadron colliders.
If the UED model is extended including non-vanishing fermion bulk mass $\mu$,
called the split UED model~\cite{sUED},
the second KK gauge boson has tree level vertices with the SM fermions.
The search for high-$p_T$ lepton plus large missing transverse energy
by the CMS experiment,
which can be explained by $pp\to W^{(2)}\to \ell\nu$, 
has set quite strong exclusion limit on $R^{-1}\gsim 800\gev$
for $\mu>100\gev$~\cite{CMS:sUED}.

As a smoking gun signal of the mUED at hadron colliders,
we focus on
the  {\em second} KK gluon.
Although its major decay modes are KK-number conserving
into $q^\tw q$ and $q^\on \bar{q}^\on$,
the decay into
SM particles is also allowed by its even parity~\cite{discrimination,KK2:Song}.
This loop-induced decay,
mediated by the first KK modes of quark and gluon/Higgs boson, is indeed substantial
since the nearly degenerate KK mass spectrum suppresses the kinematic space
of the KK-number conserving decay.
In addition, large Yukawa coupling of the top quark
enhances branching ratio of
$\gtw \to \ttb$.
This $\gtw$ is a very good candidate for the $\ttb$ resonance.
As shall be shown, the recent data at the Tevatron
have set a significant direct bound on the mUED model.
This is our main result.

The organization of the paper is as follows.
In the next section, we briefly review the model and discuss the characteristic features of
the loop-induced vertex of $\gtw$ with the SM particles.
Section \ref{sec:ttbar} deals with the production  
of the second KK gluon, followed by its decay into $\ttb$.
The current data from the search for resonant $\ttb$ production
by the CDF, D0, ATLAS,
and CMS experiments are to be analyzed to constrain the model.
Future prospects, especially through dijet resonance at the LHC,
shall be discussed also.
We conclude in Sec.~\ref{sec:conclusions}.

\section{The loop-induced vertex of $\gtw$ in the mUED model}
\label{sec:review}

%\key{model description}
The UED model is based on a flat 5D spacetime with the metric of
\bea
g_{MN} = 
\left(
\begin{array}{cc}
g_{\mu\nu} & 0 \\
0 & -1 \\
\end{array}
\right),
\eea
where $M,N=0,1,\cdots,4$,
and $g_{\mu\nu} = {\rm diag} (1,-1,-1,-1)$
is the four-dimensional (4D) metric.
The word {\em universal} is from the setup that the whole 5D spacetime
is accessible to all the SM fields.
Each SM particle has an infinite tower of KK modes. 
Chiral SM fermions from vector-like 5D fermions are 
achieved by the compactification of
the extra dimension on an $S^1/Z_2$ orbifold:
the zero mode fermion with wrong chirality is removed by imposing odd 
parity under the orbifold projection $y \to -y$,
called the $Z_2$ parity.
The radius of $S^1$ is $R$.
%In addition,
%the geometrical invariance under the exchange of two fixed points
%lead to the KK parity invariance.
The detailed expressions for the 
KK expansion of the SM field are referred to Ref. 
\cite{mued:calchep,KK2:Song}.

%\key{KK mass: tree and radiative correction}
The KK mass is of geometrical origin, which is
at tree level
\bea
\label{eq:tree:KK:mass}
M_{KK}^{(n)} =
\left\{
\begin{array}{ll}
\sqrt{  M_n^2 +  m_0^2} & \qquad\hbox{ (Boson); }\\
M_n + m_0 & \qquad\hbox{ (Fermion),}
\end{array}
\right.
\eea
where $M_n=nR^{-1}$, $n$ is called the KK number, and $m_0$ is the corresponding
SM particle mass.
Since $R^{-1} \gg m_0$, the KK mass spectrum
with the given $n$ is generically degenerate.
The radiative corrections to the KK masses~\cite{rad-correction}
play a crucial role in the phenomenologies,
determining whether a specific decay mode is kinematically allowed or not.  
In the mUED model where
boundary 
kinetic terms vanish  at the cutoff scale $\Lm$,
the radiative corrections to the KK masses are well-defined and finite.

%\key{KK number violating couplings}
A heavy KK mode decays.
At tree level, the decay respects the conservation of
KK number such as  $\gtw\to q q^\tw$
and $\gtw \to q^\on \bar{q}^\on$.
However, the high degeneracy in the KK mass spectrum
suppresses these decay modes because of small kinematic space.
For example the second KK gluon mass
for $R^{-1}=300\gev$ and $\Lm R =20$ is about 698 GeV 
while
the first KK light quark mass is about 344 GeV.
Kaluza-Klein number violating decays,
which occur at one loop level,
can be significant.
Note that KK parity $(-1)^n$ is still conserved at loop level.

%\key{Some KK-2 decay modes are prohibited}
For the phenomenological signatures of the second KK quark and gluon,
it is important to note that
some loop-induced
KK-parity conserving couplings are forbidden or negligible:
\begin{enumerate}
\item The vertex of $q^\tw$-$q$-$g$ is negligible.
The four dimensional operator 
$\bar{q}^\tw \gm^\mu \frac{\lm^a}{2} q g^{a}_\mu$ vanishes
because of the gauge invariance.
The next higher dimensional operator
$\bar{q}^\tw \sg^{\mu\nu} \frac{\lm^a}{2} q F^{a}_{\mu\nu}$
is suppressed by $1/\Lm$.
\item The vertex $g$-$g$-$\gtw$ is suppressed since it appears through 
dimension-6 operators~\cite{6dim}\footnote{We thank Ayres Freitas for pointing this out.}.
In addition,  the couplings do not have the logarithmic enhancement factor
which is about $4.6\sim 6.4$ for $\Lm R=20,50$.
Dimension-4 operators that would generate the vertex vanish by the unbroken
4D gauge invariance and the absence of the kinetic and mass mixing between $g$ and $\gtw$. 
\end{enumerate}
In what follows, therefore, we ignore the above two kinds of vertices.

The loop-induced vertex of $\gtw$
is only with the SM quarks, given by
\bea
\label{eq:Lg}
&&-i \frac{g_s}{\sqrt{2}}
\sum_{X=L,R}
\left(
\frac{ \bar{\delta} m^2_{g_2} }{M_2^2}
-
2 \frac{ \bar{\delta} m_{q_{X2}}}{M_2}
\right)
\bar{f}_X \gm^\mu  \frac{\lm^a}{2} P_X f_X g^{a\tw}_\mu
\\ \no
&\equiv&
-i \frac{g_s}{\sqrt{2}}
\left(
\frac{1}{16\pi^2}\ln \frac{\Lm^2}{Q^2}
\right)
\bar{f} \gm^\mu  \frac{\lm^a}{2} 
\left\{
\widehat{g}_{fL}P_L + \widehat{g}_{fR} P_R
\right\} 
f g^{a\tw}_\mu
,
\eea
where $P_{R,L} = (1\pm \gm^5)/2$,
$Q$ is the regularization scale,
and $\bar{\delta} m$ is the boundary mass correction~\cite{rad-correction}.
For the $\gtw$ production, 
we adopt $Q=2R^{-1}$.
The effective couplings with $\gtw$
of $t_{L,R}$ and $b_{L,R}$ are 
\bea
\label{eq:ghat}
\hat{g}_{tL}&=&
\frac{1}{8}
\left[
44 g_3^2 -27 g_2^2- g_1^2 + 12 h_t^2
\right] 
\approx 6.2,
\\ \no
\hat{g}_{bL} &=& \frac{1}{8}
\left[
44 g_3^2 -27 g_2^2- g_1^2 
\right] 
\approx 4.7,
\\ \no
\hat{g}_{tR}&=&
\frac{11}{2} g_3^2 - 2 g_1^2 + 3 h_t^2
\approx 8.9,
\\ \no
\hat{g}_{bR} &=& \frac{11}{2} g_3^2 - \frac{1}{2} g_1^2
\approx 6.1.
\eea
The couplings of $\gtw$ with light up-type (down-type)
quarks are the same as $\hat{g}_{tL,tR}$ ($\hat{g}_{bL,bR}$)
except for the top Yukawa coupling.
As explicitly shown in Eq.~(\ref{eq:ghat}),
the strong coupling of the gluon and 
the large Yukawa coupling of the top quark
play in the same direction to increase 
the branching ratio  such that
$\br(\gtw \to \ttb) \approx 7\%-8\%$,
depending on the model parameters.

\section{Direct bounds from the $\ttb$ resonance search}
\label{sec:ttbar}
%\key{Numerical results for the ratios}

\begin{figure}[t!]
\centering
\includegraphics[width=7cm]{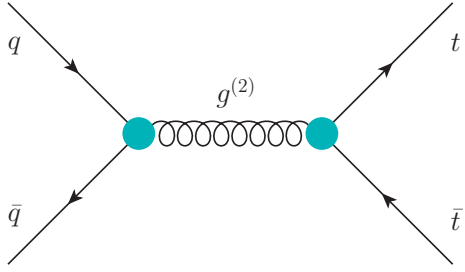}
\caption{\label{fig:feyn}\small
Feynman diagram for $\gtw$ production
as a $\ttb$ resonance at a hadron collider.
A light bulb denotes the loop-induced vertex of $\gtw$-$q$-$\bar{q}$
with the first KK modes running in the loop.}
\end{figure}

The production of the second KK gluon is through $q\bar{q}$ annihilation.
Even without large Yukawa couplings,
light quarks also have sizable couplings with $\gtw$, 
as can be seen from $\hat{g}_{bL, bR}$ in Eq.~(\ref{eq:Lg}).
The best channel to probe second KK modes at a hadron collider
is the $q\bar{q}$ annihilation production of $\gtw$, followed by the decay
$\gtw \to\ttb$: see Fig.~\ref{fig:feyn}.
Although other second KK gauge bosons like $Z^{(2)}$ and $\gm^{(2)}$
also produce $\ttb$ resonance signal,
their electroweak production
is much smaller than the $\gtw$ production
by an order of magnitude~\cite{discrimination}.

The production of $\gtw$
has additional production channels associated with soft jets.
At a hadron collider, the number of jets are measured in terms of jet multiplicity.
If jets are very soft, however,
they are very likely to be missed in
the jet multiplicity.
The heavy mass of $\gtw$ and the steeply falling parton luminosities
result in $\gtw$ production near the threshold.
The accompanying SM jets are generically soft.
Soft jets tend to spread out.
If the transverse momentum of soft jets are too low like below 20 GeV,
soft jets cannot excite showers in the hadron calorimeter of the detector.
Finally the jets with $|\eta_j|>2.5$, going out of the barrel and end caps of the hadron calorimeter,
are also missed.
Therefore, we include the $\gtw$ production with soft jets,
such as
$q\bar{q} \to  g g^\tw$, $g q \to q  g^\tw$, and $g g \to q \bar{q} \gtw$.
For the soft jets, we apply $p_T^{j}<20\gev$ or $|\eta_j|>2.5$.

\begin{figure}[t!]
\centering
\includegraphics[width=12cm]{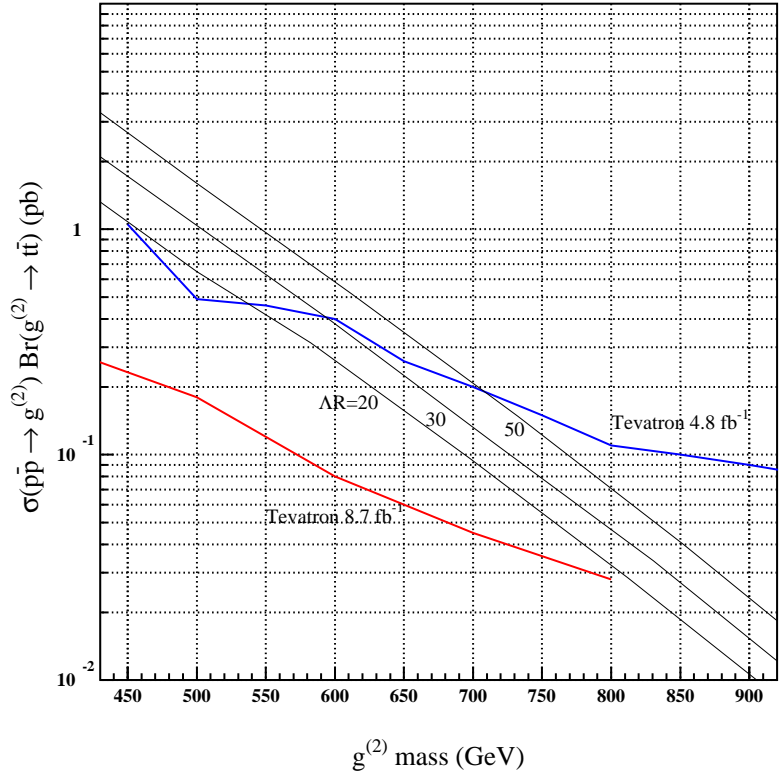}
\caption{\label{fig:Tevatron}\small
Expected and observed 95\% C.L. upper limit on
$\sigma(p \bar{p} \to \gtw \to \ttb)$
as a function of $\ttb$ invariant mass at the Tevatron with $\sqrt{s}=1.96\tev$.
}
\end{figure}

We first present the expected signal and the observed upper bound at the Tevatron.
Recently the Tevatron has improved
the $\ttb$ resonance search sensitivity by including all hadronic 
$\ttb$ decays.
The previous search was based on the final states of the lepton plus
jets~\cite{CDF:ttb:semileptonic,D0:ttb:semileptonic}.
All hadronic decay modes of $\ttb$ have 
the advantages of larger branching ratio of $W$'s hadronic decay
and the improved resolution of the invariant mass of $\ttb$
due to the absence of neutrinos.

In Fig.~\ref{fig:Tevatron}, we show the production cross section of $\gtw$
multiplied by $\br(\gtw\to \ttb)$, as a function of $\gtw$ mass
for $\Lm R=20,~30,~50$.
For the signal event generator, we have used CalcHEP~\cite{calchep}. 
Larger $\Lm R$ yields larger signal, since the loop-induced vertices increase
logarithmically with $\Lm R$, as can be seen in Eq.~(\ref{eq:Lg}).
The Tevatron search for $\ttb$ resonance based on the $4.8\fbi$ data~\cite{Tevatron:4.8}
has already set the lower bounds on $\mgtw \gsim  535\gev$ for $\Lm R=20$,
$\mgtw \gsim 590\gev$ for $\Lm R=30$,
and $\mgtw \gsim  710\gev$ for $\Lm R=50$.
The most recent data with total luminosity of $8.7\fbi$~\cite{Tevatron:8.7} extend
the exclusion region up to $\mgtw=800\gev$.
Irrespective to the value of $\Lm R$,
the signal in the mUED model exceeds the observed 95\% C.L. upper bound.
The Tevatron group presented their analysis only up to the $\ttb$ invariant mass of $800\gev$.
If naively extrapolating the observation,
we have   $\mgtw \gsim 820\gev$ for $\Lm R=20$,
$\mgtw \gsim 870\gev$ for $\Lm R=30$,
and $\mgtw\gsim 920\gev$ for $\Lm R=50$.
These are very significant direct bounds on $\mgtw$ and thus the mUED model.

\begin{figure}[t!]
\centering
\includegraphics[width=12cm]{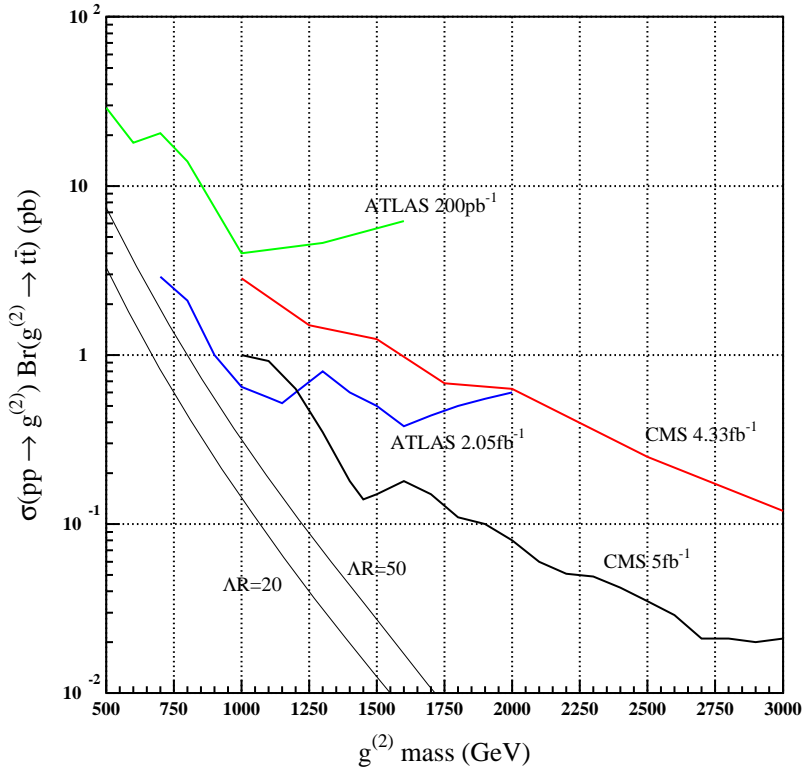}
\caption{\label{fig:LHC}\small
Expected and observed 95\% C.L. upper limit on
$\sigma(p p \to \gtw \to \ttb)$
as a function of $\ttb$ invariant mass
at the LHC with $\sqrt{s}=7\tev$.
Observed limits include the recent results by ATLAS and CMS experiment.
}
\end{figure}

Figure \ref{fig:LHC} shows the expected signal in the mUED model
and the observed 95\% C.L. upper bound at the LHC with $\sqrt{s}=7\tev$.
We present the $200\pbi$~\cite{ATLAS:ttb} and 
$2.05\fbi$ data~\cite{ATLAS:2.05} by the ATLAS experiment,
and $4.6\fbi$~\cite{CMS:ttb} and $5.0\fbi$ data~\cite{CMS:5} by the CMS experiment.
Since we have included only the $q\bar{q}$ annihilation production of $\gtw$,
this is a conservative limit.
The gluon fusion production, which is from dimension-6 operators,
is expected small and thus neglected.
Computing these operators from finite one-loop contributions 
is beyond the scope of this work.
The signal is still quite below the upper bound set by the ATLAS and CMS experiments.
Nevertheless we are still optimistic that the excellent performance of the LHC
with high luminosity will eventually cover a large portion of the model parameter space,
especially high $\mgtw$ region.

\begin{figure}[t!]
\centering
\includegraphics[scale=0.9]{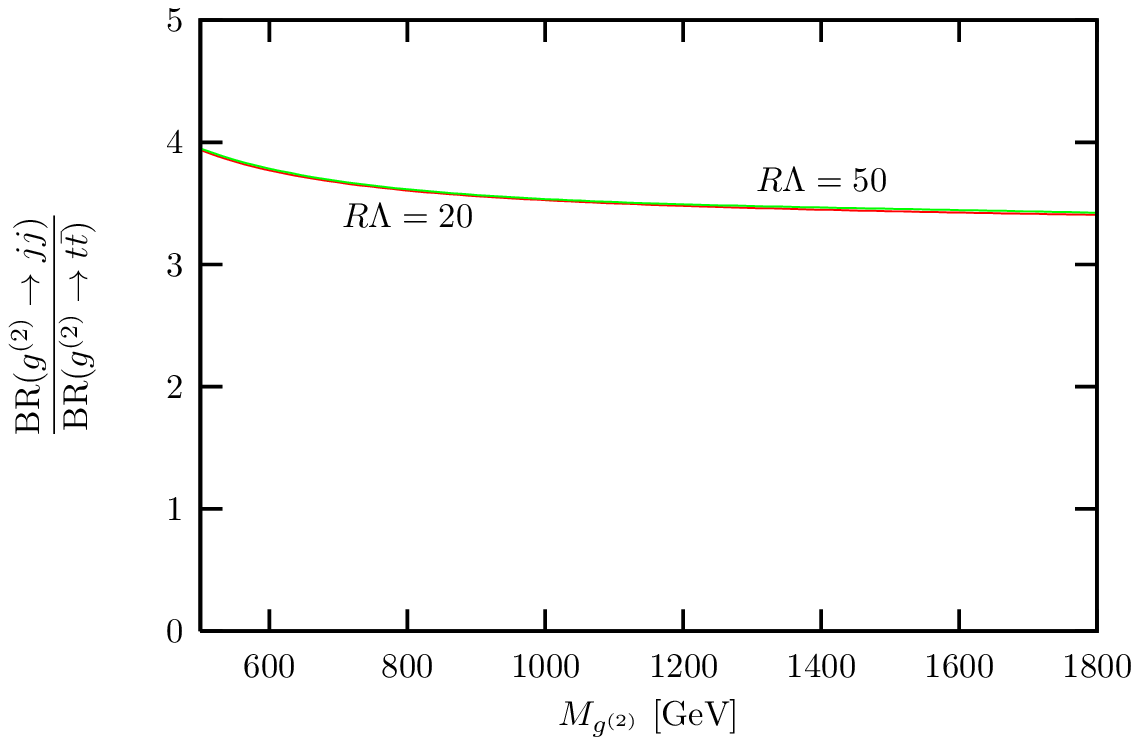}
\caption{\label{fig:BRratio}\small
The ratio $\sum_{q=u,d,c,s,b}\br(\gtw\to q\bar{q})/\br(\gtw\to \ttb)$
as a function of the $\gtw$ mass for $\Lm R=20,50$.
}
\end{figure}

%\key{Ratio}
Finally we notice one of unique features of $\gtw$ in the mUED model,
and suggest an optimal observable.
As can be seen from the effective vertex of $\gtw$-$q$-$\bar{q}$ in Eq.~(\ref{eq:Lg}),
the dependence of the model parameters $(R^{-1}, \Lm R)$ on the vertex
is common for all SM quarks.
The ratio of the $\gtw$ decay rate into $\ttb$
to that into $q\bar{q}$ is almost fixed by the SM parameters,
independent of the model parameters.
Minor dependence exists
through the mass of $\gtw$, which affects kinematics.

In Fig.~\ref{fig:BRratio}, we show the ratio $\br(\gtw\to j j)/\br(\gtw\to \ttb)$
as a function of $\mgtw$.
Two lines for $\Lm R=20,50$ 
are almost identical, as expected from the common dependence of model parameters.
In addition the value of this ratio is large about 3.5.
This is attributed to the number of flavors
although light quarks have smaller effective couplings with $\gtw$ than the top quark
because of their small Yukawa couplings.
If $\gtw$ is observed as a $\ttb$ resonance,
we should see the same resonance in dijet channel
with about 3.5 times larger rate.
This is one of the most powerful signals of the mUED model.
Recently the CMS and ATLAS experiments have reported their search for dijet resonance~\cite{dijet}.
The current upper bounds are too weak to constrain the mUED model yet.
With large data set, 
the future prospect at the LHC through the correlation between the $\ttb$
and dijet channels is very promising.

\section{Conclusions}
\label{sec:conclusions}
Despite of its various theoretical virtues,
the minimal Universal Extra Dimension (mUED) model
is one of the most elusive models to directly probe at a high energy collider.
The near-degeneracy of the Kaluza-Klein (KK) mass spectrum
buries the signals of the first KK modes,
each of which decays into a very soft SM particle with missing transverse energy.
Huge QCD backgrounds overwhelm the signals.

Turning our attention to the second KK mode, 
we have novel signatures of high mass resonances decaying into SM particles.
This kind of vertex is radiatively generated with the first KK modes running in the loop.
Since the KK-number conserving decay at tree level
is kinematically reduced,
loop-induced decay into the SM particles is enhanced.
The near-degeneracy,
which obscures the first KK mode signals,
clears the second KK mode signals.

One of the golden modes to probe the mUED model
is $p\bar{p} \to q\bar{q} \to \gtw \to \ttb$.
Strong coupling of a gluon and large Yukawa coupling of a top quark
play in the same direction to enhance the branching ratio of $\gtw\to\ttb$.
The vertex $g$-$g$-$\gtw$ appears from dimension-6 operators,
which leads to the main $\gtw$ production 
through $q\bar{q}$ annihilation.
The $p\bar{p}$ collider can be more efficient.

We have shown that the recent Tevatron $\ttb$ search 
with $8.7\fbi$ data has set very significant direct bound on the mUED model.
The $\gtw$ mass below 800 GeV is excluded for all model parameters.
If $\Lm R=50$, the lower bound is raised to about 920 GeV.
At the LHC, the absence of the gluon fusion production of $\gtw$ 
reduces the sensitivity.
No direct bounds have been derived yet.
However, the suggested correlation between the $\ttb$ resonance 
and the dijet resonance is expected to enhance the sensitivity to probe the model.

\acknowledgments
The work of JS, SS, and SC is supported partially by WCU program through the KOSEF funded
by the MEST (R31-2008-000-10057-0) and partially by
the National Research Foundation of Korea (NRF) 
funded by the Korean Ministry of
Education, Science and Technology (2011-0029758).
The work of KYL is also supported by
the Basic Science Research Program 
through the National Research Foundation of Korea (NRF) 
funded by the Korean Ministry of
Education, Science and Technology (2010-0010916).

%-------------- References ---------------------
\def\PRD #1 #2 #3 {Phys. Rev. D {\bf#1},\ #2 (#3)}
\def\PRL #1 #2 #3 {Phys. Rev. Lett. {\bf#1},\ #2 (#3)}
\def\PLB #1 #2 #3 {Phys. Lett. B {\bf#1},\ #2 (#3)}
\def\NPB #1 #2 #3 {Nucl. Phys. B {\bf #1},\ #2 (#3)}
\def\ZPC #1 #2 #3 {Z. Phys. C {\bf#1},\ #2 (#3)}
\def\EPJ #1 #2 #3 {Euro. Phys. J. C {\bf#1},\ #2 (#3)} 
\def\JHEP #1 #2 #3 {JHEP {\bf#1},\ #2 (#3)}
\def\IJMP #1 #2 #3 {Int. J. Mod. Phys. A {\bf#1},\ #2 (#3)}
\def\MPL #1 #2 #3 {Mod. Phys. Lett. A {\bf#1},\ #2 (#3)}
\def\PTP #1 #2 #3 {Prog. Theor. Phys. {\bf#1},\ #2 (#3)}
\def\PR #1 #2 #3 {Phys. Rep. {\bf#1},\ #2 (#3)}
\def\RMP #1 #2 #3 {Rev. Mod. Phys. {\bf#1},\ #2 (#3)}
\def\PRold #1 #2 #3 {Phys. Rev. {\bf#1},\ #2 (#3)}
\def\IBID #1 #2 #3 {{\it ibid.} {\bf#1},\ #2 (#3)}

\end{document}